\documentclass[10pt,letterpaper]{article}
\usepackage{opex3}
\usepackage{cite}
\usepackage{graphicx}
\usepackage{amsmath}
\usepackage{amsfonts}
\usepackage{bm}

\begin{document}

\title{Compensating the electron beam energy spread by the natural transverse gradient of laser undulator in all-optical x-ray light sources}

\author{Tong~Zhang, Chao~Feng, Haixiao~Deng,$^*$ Dong~Wang, Zhimin~Dai and Zhentang~Zhao}
\address{Shanghai Institute of Applied Physics, Chinese Academy of Sciences, Shanghai 201800, China}
\email{$^*$denghaixiao@sinap.ac.cn}

\begin{abstract}
All-optical ideas provide a potential to dramatically cut off the size and cost of x-ray light sources to the university-laboratory scale, with the combination of the laser-plasma accelerator and the laser undulator. However, the large longitudinal energy spread of the electron beam from laser-plasma accelerator may hinder the way to high brightness of these all-optical light sources. In this paper, the beam energy spread effect is proposed to be significantly compensated by the natural transverse gradient of a laser undulator when properly transverse-dispersing the electron beam. Theoretical analysis and numerical simulations on conventional laser-Compton scattering sources and high-gain all-optical x-ray free-electron lasers with the electron beams from laser-plasma accelerators are presented.
\end{abstract}


X-ray continuously revolutionizes the understanding of the matters, and creates new sciences and technologies, which has been proved by $20$ Nobel Prizes awarded for x-ray related works. Therefore, synchrotron radiation light sources and free-electron lasers (FELs) based on particle accelerators are being developed worldwide to satisfy the dramatically growing demands in the material and biological sciences~\cite{McNeil_2010_XFEL_NP}. Recently, the successful operation of the first FEL facilities in the XUV and hard x-ray region announced the birth of the coherent x-ray science era~\cite{Ackermann_2007_FLASH_NP,Allaria_2012_FermiFEL_NP,Emma_2010_LCLS_NP,Ishikawa_2012_SACLA_NP}. However, considering the size and cost of these large-scale facilities, scientists begin to envision much more compact x-ray light sources~\cite{Schoenlein_1996_ThomsonScattering_Science,Schlenvoigt_2007_LPASR_NP,Gruner_2007_tabletopXFEL_APB,Kazuhisa_2008_tableTopXFEL_NP,Fuchs_2009_LPAsource_NP,Sprangle_2009_AOFEL_PRSTAB,Kneip_2010_tabletop_NP,Deng_2012_HarmonicXFELO_PRL} which may be popularized and afforded by the university-scale laboratories.

The marriage between the advanced laser technique and accelerators has contributed significantly to the miniaturization of the x-ray sources. On one hand, the developing of optical undulator shortens the period length towards down to sub-centimeter level~\cite{Shintake_1983_MU_JJAP,Chang_2012_LU_APL}, and thus reduces the electron beam energy required by the x-ray light sources from several GeV to tens of MeV, e.g. the well-known laser-Compton scattering light sources~\cite{Schoenlein_1996_ThomsonScattering_Science,Sprangle_2009_AOFEL_PRSTAB,Kashiwagi_2009_EUVcompton_RPC,Krafft_2010_ComptonSource_RAST,Ta_2012_compton_NP,Chen_2013_comptonScattering_PRL}. On the other hand, the laser-plasma accelerators (LPAs) have made a great breakthrough in the generation of electron beam with peak current above $10\,\mathrm{kA}$, low emittance less than $0.1\,\mathrm{{\mu}m}$ and beam energy of $1\,\mathrm{GeV}$~\cite{Leemans_2006_GeVLPA_NP,Lundh_2011_LPA_NP,Fritzler_2004_LPAemittance_PRL,Brunetti_2010_LPAemittance_PRL,Liu_2011_LPA_PRL,Plateau_2012_LPAemittance_PRL}. Taking the great advantages of the laser undulator and the LPA, novel table-top all-optical Compton gamma-ray source has already been experimentally demonstrated~\cite{Ta_2012_compton_NP}. And more recently, high-gain all-optical x-ray FELs are being under consideration~\cite{Petrillo_2008_AOFEL_PRSTAB,Chang_2013_HGXFEL_PRL,Chang_2013_HGXFEL_OE}.

Currently, the electron beam from LPA usually has large energy spread with a few percent level, which may be the bottleneck for achieving high brightness in the x-ray light source applications. Thus, various great efforts have been made to reduce the energy spread in the LPA community~\cite{Mangles_2004_LPAper_Nature,Geddes_2004_LPAchannel_Nature,Leemans_2006_GeVLPA_NP}. Meanwhile, in the light source community, the concept of transverse gradient undulator (TGU) proposed in FEL oscillator~\cite{Smith_1979_TGU_JAP} has been recently suggested to overcome the electron beam energy spread in the LPA driving high-gain FELs~\cite{Huang_2012_TGU_PRL}, ultimate storage rings~\cite{Cai_2013_TGUpepx_SLACPUB} and seeded FEL regime~\cite{Deng_2013_PEHG_PRL,Feng_2014_PEHG_NJP}. Novel design of TGU has been also used in superconducting undulator to achieve much higher field gradient~\cite{Fuchert_2012_novelUndu_NIMA}. In this paper, we explore a novel concept for compensating the beam energy spread effect in the all-optical x-ray sources by using the nonlinear Gaussian distribution of laser, i.e., the controllable natural transverse field gradient of the laser undulator (TGLU). Theoretical analysis and three-dimensional simulations indicate that the radiation pulses from TGLU boosted x-ray sources present orders of magnitude brightness enhancement when compared with normal all-optical sources driven by the LPA beams with relatively large energy spread, as well as the significantly improved spatial and temporal characteristics.

\begin{figure}[!t]
	\centering
	\includegraphics[width=0.8\linewidth]{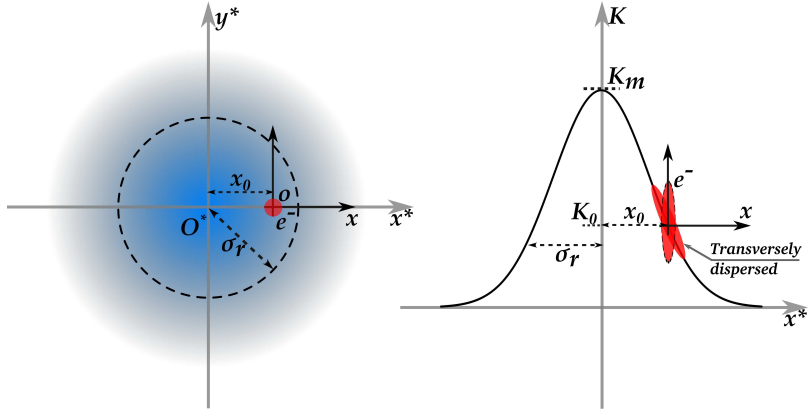}
	\caption{\emph{Left}: The electric field distribution of laser undulator and the incident electron beam in the transverse plane, where $x^*o^*y^*$ and $xoy$ is the coordinates of the laser undulator and the electron beam, respectively, $\sigma_r$ is rms transverse radius of the laser undulator, $x_0$ is the centroid displacement between the electron beam and the laser undulator, '$+/-$' sign of $x_0$ indicates that the electron beam is located at the right or left side of the origin point $o^*$, the longitudinal coordinates $z^*$ and $z$ are perpendicular to the transverse plane. \emph{Right}: the dimensionless strength parameter of the laser undulator with transverse nonlinear Gaussian gradient, and the $\left(x,\gamma\right)$ phase-space of the electron beam with/without transverse dispersion.}
    \label{fig:TGLUprinciple}
\end{figure}

In order to generate table-top x-ray sources, all-optical scheme consists of the LPA and the laser undulator has been proposed. The principle of the laser undulator is more generally known as laser-Compton scattering, where the electrons with relativistic energy $\gamma_0$ scatter the laser photons with wavelength $\lambda_l$ to a shorter wavelength $\lambda_s$ at the expense of the electrons kinetic energy~\cite{Compton_1923_ComptonScattering_PR}. The interaction between the electron and the laser undulator can be described by the undulator resonant relation from the conventional theory of undulator radiation, i.e.,
\begin{equation}\label{eqn:resonant}
\lambda_s = \frac{\lambda_u}{2\gamma_0^2} \left( 1+\frac{K_0^2}{2} \right)
\end{equation}

The equivalent period length of the laser undulator $\lambda_u$ is determined by the scattering angle $\phi$ of the electron beam and the laser by $\lambda_u = \lambda_l/\left(1-\cos \phi \right)$, the dimensionless undulator parameter $K$ could be written as $K = e E \lambda_u/\left(2\pi m c^2\right)$, where $e$, $E$, $m$ and $c$ is the elementary charge, the amplitude of laser field, the rest electron mass and the speed of light in vacuum, respectively. It is expected that $K$ could reach $1-2$ with the state-of-the-art laser technique~\cite{Du_1994_Laser_APL,Chang_2013_HGXFEL_PRL}, and still be moving forward.

To clearly illustrate the transverse gradient of a laser undulator, we start from the case where the electron beam interacts with a counter-propagating laser undulator, i.e., $\phi=\pi$. The transverse field of the laser undulator can be seen in Fig.~\ref{fig:TGLUprinciple}., into which the electron beam is guided with a transverse displacement of $x_0$, thus in the frame of electron beam, the undulator parameter felt by the electron at the position of $x$ could be written as
\begin{equation}\label{eqn:kx}
  K\left(x\right)=K_0 e^{-\frac{\left(x+x_0 \right)^2 - x_0^2}{2\sigma_r^2}}
\end{equation}

where $K_0=K_m e^{-\frac{x_0^2}{2\sigma_r^2}}$ is the undulator parameter at the center of electron beam. Then the gradient of the laser undulator can be observed from the derivative of Equation~\eqref{eqn:kx}, i.e.,
\begin{equation}\label{eqn:kxprime}
  \frac{\Delta K}{K} = -\frac{x_0}{\sigma_r^2} \Delta x
\end{equation}

Thus, the equivalent transverse field gradient of laser undulator could be written as $\alpha=-\frac{x_0}{\sigma_r^2}$, which can be controlled obviously by adjusting the laser beam size and the injection position of the electron beam. Under such circumstance, if the electron beam with large energy spread was properly dispersed by $x=\eta \Delta \gamma/\gamma$ in the transverse plane, with $\eta$ the dispersion strength, the longitudinal energy spread effects of the electron beam could be converted to the transverse plane and could be potentially compensated.

\begin{figure}
	\centering
	\includegraphics[width=0.6\linewidth]{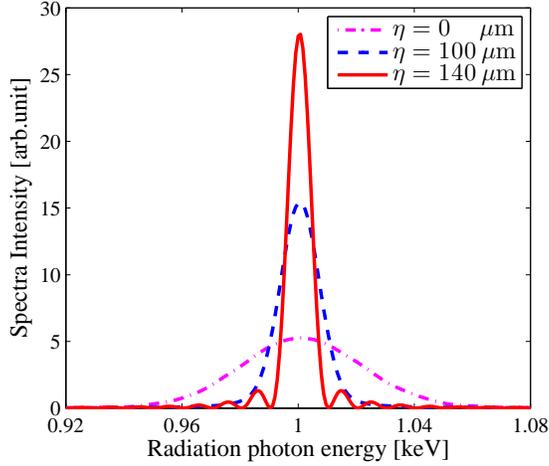}
	\caption{Calculated spectra of laser-Compton scattering with different strengths of transverse dispersion. The energy of the cold beam is $45\,\mathrm{MeV}$ with energy spread of $1\%$, and the beam size of laser undulator is $\sigma_r=100\,\mathrm{{\mu}m}$, displacement between the laser and electron is $x_0=-100\,\mathrm{{\mu}m}$.}
    \label{fig:TGLUlowgain}
\end{figure}

The essence of the TGLU on suppressing the beam energy spread effects can be learnt from the laser-Compton scattering process. Here we consider a $10\,\mathrm{{\mu}m}$ wavelength laser with $100\,\mathrm{{\mu}m}$ radius and $3.3\,\mathrm{ps}$ flat-top distribution in longitudinal, i.e., a laser undulator with period number $N_u=100$, and a cold electron beam with energy of $45\,\mathrm{MeV}$ and relative energy spread of $1\%$, then an effective undulator strength $K_0=2.3$ results in radiation of $1\,\mathrm{keV}$ photon energy. According to the theory of undulator radiation, the bandwidth of $1\,\mathrm{keV}$ photon will be $1\%$ for cold electron beam, which however will be significantly broadened by a large beam energy spread. If the electron beam with a dispersion $\eta$ was injected at $x_0=-100\,\mathrm{{\mu}m}$, the radiation spectrum on the beam axis can be calculated from the sum of the spontaneous emission spectrum of individual electrons. Figure~\ref{fig:TGLUlowgain}.~shows the radiation spectra with different dispersions to the electron beam, it is found that the radiation bandwidth is approximately $6$ times improved with $\eta=140\,\mathrm{{\mu}m}$, when compared with the case of $\eta=0\,\mathrm{{\mu}m}$, which in fact already achieves the theoretical limit of the undulator radiation, i.e. the energy spread effect has been successfully compensated.

It is well known that the FEL gain at small-signal regime is proportional to the derivative of the undulator spontaneous emission spectra, which greatly depends on the beam energy detuning, i.e., $\Delta \gamma/\gamma_0$~\cite{Madey_1971_FEL_JAP}. In order to maximize the small-signal gain, in principle, $\Delta \gamma/\gamma_0$ should be kept as a constant with different $x$ value. Then theoretically one can derive the optimal relationship between the TGLU and the dispersion as follows~\cite{Huang_2012_TGU_PRL},
\begin{equation}\label{eqn:etaAlpha}
  -\frac{x_0}{\sigma_r^2} = \frac{2+K_0^2}{\eta K_0^2}
\end{equation}

In the cases of Fig.~\ref{fig:TGLUlowgain}., the gradient of the laser undulator is $10^5\,\mathrm{m^{-1}}$ and the Eq.~\eqref{eqn:etaAlpha} predicts the optimal dispersion of $140\,\mathrm{{\mu}m}$, which is pretty consistent with the simulated results. Moreover, the spectra also demonstrate a maximum small-signal gain for the case of $140\,\mathrm{{\mu}m}$ dispersion strength.

Recently, it is reported that the powerful x-ray from high-gain Compton-scattering could be realizable by significantly extending the interaction length of the electron beam and the laser undulator, but rather small beam energy spread of $5\times10^{-5}$ was critically required~\cite{Chang_2013_HGXFEL_PRL}. Here we demonstrate the advantages of TGLU in the application of high-gain all-optical x-ray FELs, where the collision angle is $\phi=\pi/2$, then accordingly the transverse laser shape should be subtly manipulated to introduce the proper transverse gradient~\cite{Chang_2013_HGXFEL_PRL}. For the transversely dispersed electron beam in the laser undulator, taking the coupling reduction between the electron beam and the radiation into accounts, the effective FEL parameter can be defined as~\cite{Huang_2012_TGU_PRL},
\begin{equation}\label{eqn:rhot}
\rho_\mathrm{T}=\rho \left(1+\frac{\eta^2\sigma_{\delta}^2}{\sigma_x^2}\right)^{-\frac{1}{6}}
\end{equation}
where $\sigma_x$ is the electron beam radius, $\sigma_\delta$ is the fractional energy spread,$\rho$ is the classical FEL parameter~\cite{Bonifacio_1984_FEL_OC}. Assuming the average beta function of the LPA beam in laser undulator to be $\bar{\beta}=N_u \lambda_u/2$, then including the transverse gradient of a laser undulator, the FEL power gain length can be written similarly as that in Ref.~\cite{Huang_2012_TGU_PRL}:
\begin{equation}\label{eqn:lgt}
L_g^\mathrm{T} \approx \frac{\lambda_u}{4\pi \sqrt{3}\rho_\mathrm{T}} \left(1+\left(\frac{K_0^2}{2+K_0^2}\right)^2 \frac{x_0^2\epsilon_n N_u \lambda_u}{2\gamma \rho_\mathrm{T}^2\sigma_r^4}\right)
\end{equation}
with $N_u$ the total period number of laser undulator.

Equation~\eqref{eqn:lgt} gives the clue about the gain length shortening for relative large beam energy spread, which means the energy spread reduction could be compensated by TGLU. For instance, the LPA beam with energy spread of $1\%$ and other parameters shown in Table~\ref{tab:TGLU_highgain}, the laser undulator with $x_0=\sigma_r=100\,\mathrm{{\mu}m}$, dispersion $\eta=-140\,\mathrm{{\mu}m}$ would contribute $54\%$ shortening to the FEL gain length.

\begin{table}
    \centering
    \caption{Parameters of high-gain Compton scattering all-optical x-ray source}
    \label{tab:TGLU_highgain}
	\begin{tabular}{llll}
    \hline\hline
    Parameter & Symbol & Value & Unit \\
    \hline
    Beam Energy             & $E_b$                 & $60$      & $\mathrm{MeV}$    \\
    Relative Energy Spread  & $\sigma_\delta$       & $0.5\%$   &                   \\
    Norm. Emittance         & $\epsilon_n$          & $50$    	& $\mathrm{nm}$ 	\\
    Beam Size (rms)         & $\sigma_x$            & $1.5$     & $\mathrm{{\mu}m}$ \\
    Peak Current            & $I_p$                 & $3$       & $\mathrm{kA}$     \\
    Laser Wavelength        & $\lambda$             & $10$      & $\mathrm{{\mu}m}$ \\
	FEL Wavelength 			& $\lambda_{ph}$ 		& $1.0$ 	& $\mathrm{nm}$ 	\\
    FEL Photon Energy       & $E_\mathrm{ph}$       & $1.24$    & $\mathrm{keV}$    \\
    \hline\hline
    \end{tabular}
\end{table}

\begin{figure}[!h]
	\centering
	\includegraphics[width=0.8\linewidth]{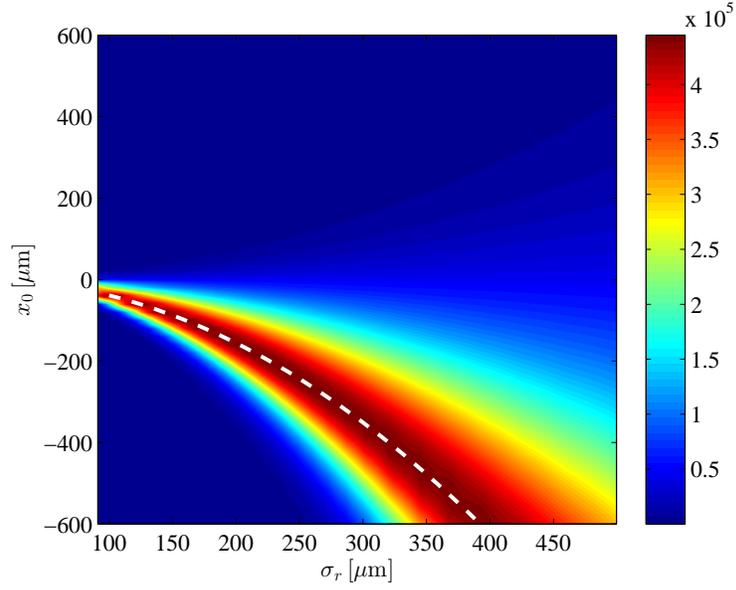}
	\caption{Radiation power contour view v.s. the transverse position $x_0$ of the incident electron beam and the transverse radius $\sigma_r$ of the laser undulator, the white dashed line indicated the optimal relation between $x_0$ and $\sigma_r$ with $x_0=-\alpha_\mathrm{opt} \sigma_r^2$, where $\alpha_\mathrm{opt}$ is optimal gradient with regard to the optimal dispersion of $\eta=480\,\mathrm{{\mu}m}$.}
    \label{fig:TGLUscan}
\end{figure}

Thanks to the continuous progresses in the LPA community, we are optimistic about achieving the beam parameters shown in Table~\ref{tab:TGLU_highgain}~\cite{Hidding_2012_LPA_PRL,Li_2013_LPA_PRL}. In order to illustrate the great potential of TGLU in the all-optical x-ray FELs, here we present the detailed three-dimensional numerical simulations. The well-benchmarked FEL code \textsc{genesis 1.3}~\cite{Reiche_1999_genesis_NIMA} is modified to adapt the TGLU configuration. After the two-dimensional scan of the transverse dispersion $\eta$ and the gradient $\alpha$ of the laser undulator against the radiation power, the optimal working condition $\eta=480\,\mathrm{{\mu}m}$ and $\alpha_\mathrm{opt}=3900\,\mathrm{m^{-1}}$ is found. Furthermore, according to the theory of TGLU, with the optimal dispersion of $480\,\mathrm{{\mu}m}$, a bundle of $\left( \sigma_r,x_0 \right)$ could be operational. The simulation results in Fig.~\ref{fig:TGLUscan}.~show the working region highlighted by the fitted curve $x_0=-\alpha_\mathrm{opt} \sigma_r^2$, e.g. $\alpha_\mathrm{opt}=3900\,\mathrm{m}^{-1}$ for $\left(\sigma_r,x_0 \right)=\left(200\,\mathrm{{\mu}m},-156\,\mathrm{{\mu}m}\right)$. It is worth to stress that the optimal working region becomes larger as the transverse radius of the laser undulator increases, which could be mathematically seen from $\left| \Delta x_0 \right|= 2 \alpha_\mathrm{opt} \sigma_r \Delta\sigma_r$.

\begin{figure}[!h]
	\centering
	\includegraphics[width=0.8\linewidth]{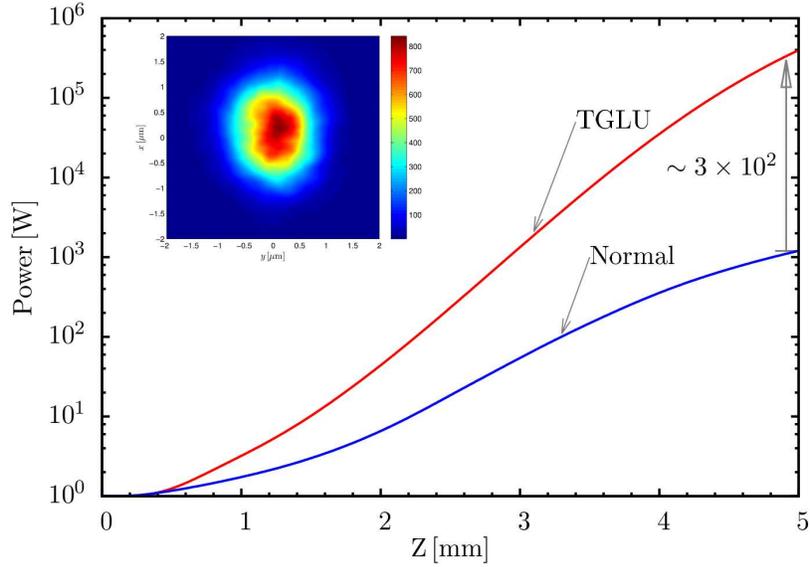}
	\caption{Power gain curve for the case of $0.5\%$ energy spread, in which the TGLU boosted case (red solid line) shows an additional gain of $300$-fold over the normal case (blue solid line); the up left inset shows the transverse field pattern for TGLU case.}
    \label{fig:TGLUgc}
\end{figure}

Figure~\ref{fig:TGLUgc}.~shows the simulated FEL power growth around $1.2\,\mathrm{keV}$ photon energy for the normal laser undulator case and the TGLU case. One can see that the power improvement of the TGLU case is almost $300$ times of magnitude within the $5\,\mathrm{mm}$ long laser-beam interaction length even though it is still far away from the FEL saturation. Further studies indicate that, for the case of $1\%$ beam energy spread, at least $1$ order of magnitude improvement of FEL power can be achieved.

Moreover, TGLU also results much better transverse radiation field pattern which could be seen from Fig.~\ref{fig:TGLUgc}. While the transverse coherence is very poor because of the large energy spread and relatively low gain with the absence of TGLU. 

Finally, we would like to discuss some practical issues about the TGLU application in the x-ray light source generation. Because the electron beam is sent into the laser undulator with $x_0$ displacement with regards to the centroid, the beam should be deflected by the non-balanced magnetic field, or the beam centroid trajectory be deviated by $\Delta x = \frac{\alpha K_0^2}{4\gamma^2}(N_u\lambda_u)^2 \approx 6\,\mathrm{\mu m}$, which would make the radiation power growth not sustainable. Well, numerical simulation shows that the external correction field with the magnitude of about $0.1$ T would be enough to correct the trajectory back into normal. Other issues like beam energy jitter and incident position offset are also substantially studied. The electron beam incident position offset error is originated both from the intrinsic random incident error and the beam energy jitter induced incident deviation, thus the latter is energy-correlated while the former is energy-uncorrelated. Numerical study shows that the random energy-uncorrelated incident offset of $\sim 5\,\mathrm{\mu m}$ in root mean square would keep the radiation power level up the case without TGLU applied. Since the energy-correlated incident offset only contributes the TGLU gradient shifting, the FEL resonant condition is always maintained, thus the requirement is not so much stringent~\cite{Smith_1979_TGU_JAP}, e.g. rms beam energy jitter of $1\%$ only slightly move the incident position of $4.8\,\mathrm{\mu m}$, which is a small portion of the optimal working region as shown in Fig.~\ref{fig:TGLUscan}. Similarly, within the rms incident angle offset of about $300\,\mathrm{\mu rad}$, the TGLU x-ray light source would be stable.

In conclusion, a novel operation mode of all-optical x-ray light source was proposed with the electron beam from the advanced laser-plasma accelerator and laser undulator. Using the natural transverse gradient of the laser undulator, the beam energy spread impact to the power gain of x-ray source could be effectively compensated. The theory and numerical simulation indicate that the optimal working condition could be found with different transverse radius of laser undulator and the electron beam incident position. The utilization of TGLU in the laser-Compton scattering shows the great advantage of almost orders of brightness enhancement by narrowing the bandwidth. The high-gain all-optical x-ray FEL light source example shows that with the laser beam size of hundreds $\mathrm{{\mu}m}$, TGLU could increase the output power by about $2$ orders of magnitude with much better transverse field mode. It is convinced that the natural transverse gradient of the laser undulator holds great potential in developing table-top all-optical x-ray light sources, and it also can be extended to the applications in the radio frequency and Tera-hertz undulators.

\section*{Acknowledgments}
The authors are grateful to C.~Chang, Y.~Ding, Z.~Huang, C.~Schroeder, L.~Shen, D.~Huang, B.~Liu and B.~Jiang for helpful discussions. This work was partially supported by the Major State Basic Research Development Program of China (2011CB808300) and the National Natural Science Foundation of China (11175240, 11205234 and 11322550).

\end{document}